\documentclass[lettersize,journal]{IEEEtran}
\usepackage{amsmath,amsfonts}
\usepackage{algorithmic}
\usepackage{algorithm}
\usepackage{array}
\usepackage[caption=false,font=normalsize,labelfont=sf,textfont=sf]{subfig}
\usepackage{textcomp}
\usepackage{stfloats}
\usepackage{url}
\usepackage{verbatim}
\usepackage{graphicx}
\usepackage{cite}
\usepackage{acronym}
\newacro{AI}{Artificial Intelligence}
\newacro{DNN}{Deep-Neural-Network}
\newacro{CNN}{Convolutional Neural Network}
\newacro{RNN}{Recurrent Neural Network}
\newacro{NN}{neural network}
\newacro{SNN}{Spiking Neural Network}
\newacro{ReRAM}{resistive RAM}
\newacro{MTJ}{magnetic tunnel junction}
\newacro{DW}{domain-wall}
\newacro{FM}{ferromagnet}
\newacro{HM}{heavy-metal}
\newacro{AFM}{antiferromagnetic}
\newacro{LIF}{Leaky Integrate-and-Fire}
\newacro{MTJ}{magnetic tunnel junction}
\newacro{LTP}{long-term potentiation}
\newacro{LTD}{long-term depression}
\newacro{TL}{top-layer}
\newacro{BL}{bottom-layer}
\newacro{MNIST}{Modiﬁed National Institute of Standards and Technology}
\newacro{PCM}{phase-change memory}
\newacro{eNVM}{emerging non-volatile memory}

\begin{document}

	\title{A Unified Evaluation Framework for Spiking Neural Network Hardware 
		Accelerators Based on Emerging Non-Volatile Memory Devices}
	
	\author{Debasis Das, Xuanyao Fong, Member, IEEE
	        % <-this % stops a space
	\thanks{The authors are with the Dept. of Elect. \& Comp. Engineering, National University of Singapore, Singapore, 117583 (e-mail: {eledd,kelvin.xy.fong}@nus.edu.sg).}}% <-this % stops a space
	%\thanks{Manuscript received April 19, 2021; revised August 16, 2021.}}
	
	\markboth{Journal of \LaTeX\ Class Files,~Vol.~XX, No.~X, XXXX~XXXX}%
	{Shell \MakeLowercase{\textit{et al.}}: A Sample Article Using IEEEtran.cls for IEEE Journals}
	
	
	\maketitle
	
	\begin{abstract}
	\acp{SNN} have emerged as a promising paradigm, offering event-driven and energy-efficient computation. In recent studies, various devices tailored for SNN synapses and neurons have been proposed, leveraging the unique characteristics of \ac{eNVM} technologies. While substantial progress has been made in exploring the capabilities of SNNs and designing dedicated hardware components, there exists a critical gap in establishing a unified approach for evaluating hardware-level metrics. Specifically, metrics such as latency, and energy consumption, are pivotal in assessing the practical viability and efficiency of the constructed neural network. In this article, we address this gap by presenting a comprehensive framework for evaluating hardware-level metrics in \acp{SNN} based on non-volatile memory devices. We systematically analyze the impact of synaptic and neuronal components on energy consumption providing a unified perspective for assessing the overall efficiency of the network. In this study, our emphasis lies on the neuron and synaptic device based on magnetic skyrmions. Nevertheless, our framework is versatile enough to encompass other emerging devices as well. Utilizing our proposed skyrmionic devices, the constructed SNN demonstrates an inference accuracy of approximately 98\% and achieves energy consumption on the order of pJ when processing the \ac{MNIST} handwritten digit dataset.
	\end{abstract}
	
	
	\section{Introduction}\label{Intro}
The field of \ac{AI} has witnessed a remarkable stride, with \acp{NN} emerging as the cornerstone of numerous applications. In computer vision, \acp{DNN} have played a pivotal role in revolutionizing image classification~\cite{ciregan2012multi}, object detection~\cite{szegedy2013deep}, and segmentation~\cite{paszke2016enet}. Models like \acp{CNN} have become instrumental in applications ranging from medical image analysis~\cite{havaei2017brain,myronenko20193d} to autonomous vehicles~\cite{kuutti2020survey}. Similarly, in natural language processing~\cite{khurana2023natural,chowdhary2020natural}, \acp{RNN}~\cite{sherstinsky2020fundamentals} and transformer architectures~\cite{devlin2018bert} have propelled machine translation, text generation, and sentiment analysis to unprecedented accuracies. With time, these \acp{NN} are expected to perform more complex and sophisticated tasks that come with inherent drawbacks, particularly in terms of hardware requirements and energy consumption. An enormous amount of energy is consumed during the training process involving iterative forward and backward passes through the network, leading to extensive matrix multiplications and weight updates~\cite{gim2022memory}. A significant amount of energy is also consumed during inference due to the high precision weights and continuous activation values. The energy further amplifies as the parameters within the neural network increase. For instance, the training of GPT-3 was estimated to consume 190,000 kWh of energy~\cite{eshraghian2023training}. An application resembling ChatGPT, with an anticipated usage of 11 million requests per hour, results in emissions of 12.8 k metric tons of $\mathrm{CO_2}$ per year—25 times higher than the emissions generated during the training of GPT-3~\cite{chien2023reducing}. Hence, an alternative is required that attains comparable accuracy while significantly reducing energy consumption. 

\ac{SNN}, draws inspiration from the architecture of the mammalian brain, where neurons communicate through spikes. In \ac{SNN}, neurons produce spikes (events) only upon reaching a specific threshold voltage, resulting in an event-driven process. Owing to this, sparse activity is observed, where most neurons remain inactive most of the time. By generating spikes only when needed, rather than continuously processing information, \ac{SNN} reduces power consumption, making them more energy-efficient compared to continuously active neural networks. There are several models to emulate the spiking neuron behavior, such as the Hodgkin-Huxley model~\cite{hodgkin1952quantitative}, Izhikevich model~\cite{izhikevich2003simple}, \ac{LIF} model~\cite{lapicque1907recherches}. Among these models, \ac{LIF} is widely used~\cite{lu2022linear,kornijcuk2016leaky,das2022design} due to its simplicity, efficiency, and ability to capture essential aspects of the neuronal behavior. In this model, an excitatory stimulus leads to an increase in the membrane potential, a process known as \textit{integration}. In the absence of any stimulus, the neuron's membrane potential gradually returns to a resting state, referred to as \textit{leak}. If, during the integration, the membrane potential surpasses a certain threshold, the neuron undergoes a \textit{firing} event, producing an output spike. Following this, the neuron enters a refractory period during which its membrane potential is reset to the resting potential, rendering the neuron unresponsive to input spikes until the refractory period concludes. Another key part of the \ac{SNN}, is synapse, associated with a weight that determines the strength of the connection between the pre-synaptic neuron and the post-synaptic neuron.

For implementing the \acp{SNN} in hardware, emerging devices such as \ac{ReRAM}, \ac{PCM}, and spintronics devices hold great promise due to their non-volatility and low-power operations~\cite{she2019improving,munoz2020hardware,sengupta2016hybrid}. 
Many reports have shown hardware implementations of the neuron and synapse operation with such devices. Gong \textit{et al.}~\cite{gong2020lateral} fabricated a two-terminal perovskite-based synapse that shows an energy consumption of 14.3~fJ per synaptic event. Yoon \textit{et al.}\cite{yoon2017synaptic} showed the energy consumption for the potentiation $\sim$22~aJ for a ferroelectric synapse. Using micromagnetic simulation, Sengupta \textit{et al.}~\cite{sengupta2017performance} estimated the energy consumption of a magnetic domain wall-based neuron is $\sim$5.7~fJ. Elkin \textit{et al.}~\cite{cruz2022evaluating} showed the energy consumption of 15~fJ/spike for the neuron and 0.081~aJ per synaptic operation. While the numerical values appear promising concerning energy efficiency, it is imperative to note that these calculations pertain exclusively to the devices themselves and do not account for the access transistors and related circuitry. Therefore, we believe it is of paramount importance to establish a benchmark for realistic energy figures that take into account the circuits associated with the neurons and synapses implemented by emerging devices.

This study introduces a comprehensive framework for assessing the energy consumption of \ac{SNN}, encompassing the incorporation of access transistors required to operate neuron and synaptic devices. While the outcomes presented here are obtained by our proposed bilayer magnetic skyrmion device implementing neurons and synapses, it is essential to note that the proposed framework also applies to other emerging devices. We evaluate the energy consumption of the \ac{SNN} by performing a classification task on the \acf{MNIST} handwritten digit dataset. In Section \ref{Sec:snn_simlation}, we discuss the network architecture of the \ac{SNN}. Next, we describe the implementation of the neuron and synapse using a bilayer magnetic skyrmion device, in brief in Section~\ref{Sec:Device}. Section~\ref{Sec:circuit}, describes the associated circuits for the operation, and finally, Section~\ref{Sec:conclusion} concludes the paper.

    \section{Framework}\label{Sec:framework}
\begin{figure}[h!]
\centering
	\includegraphics[scale=0.4]{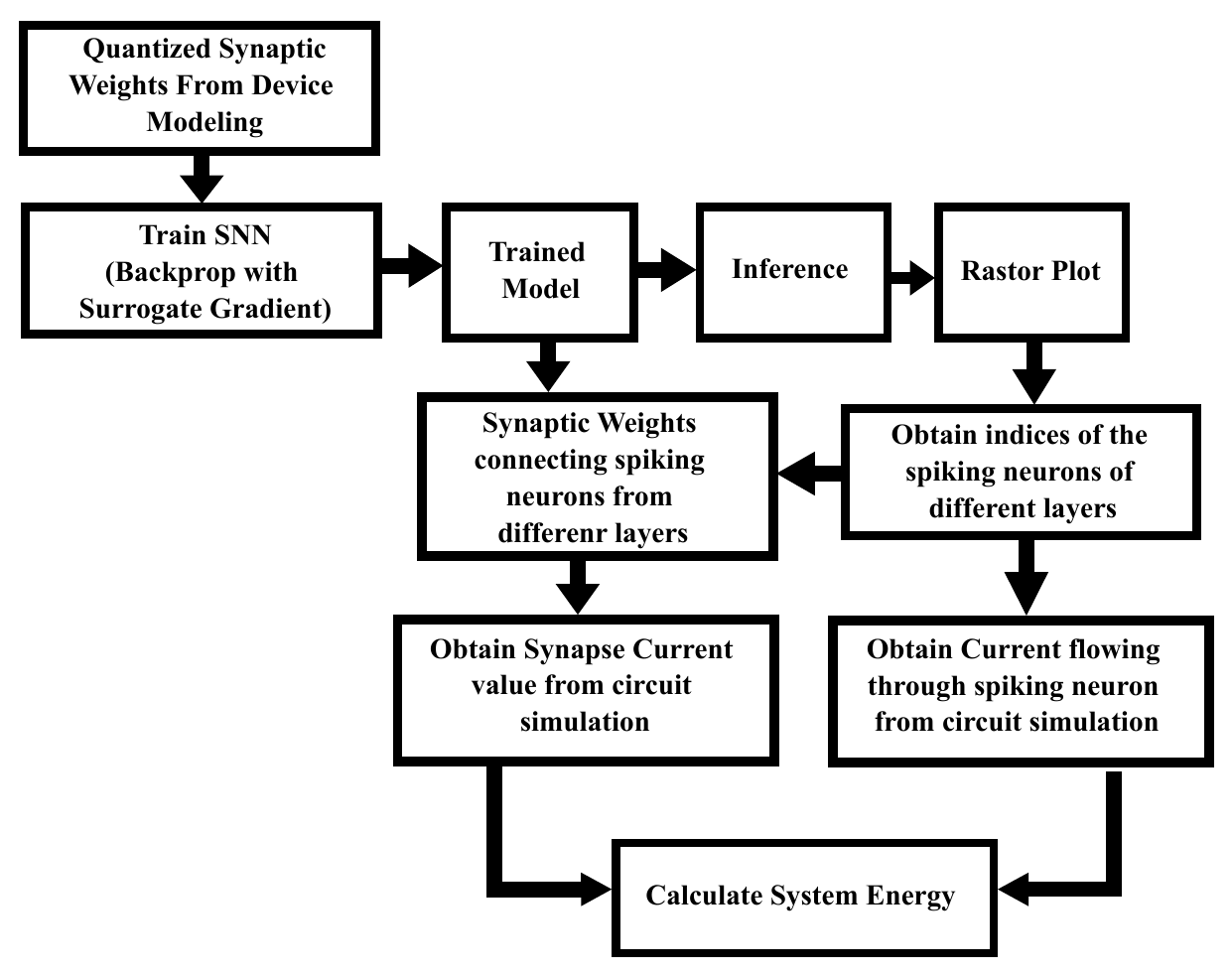}
	\caption{Overall framework for estimating energy for \ac{SNN}}
	\label{Fig:framework}
\end{figure}
Figure~\ref{Fig:framework} illustrates the overarching framework employed to estimate energy consumption for inference tasks across various datasets. This methodology enables the estimation of network energy consumption regardless of the underlying device technology utilized for neuron and synaptic devices. While full-precision weights are employed in software for neuromorphic tasks, edge devices typically utilize lower-resolution synaptic weights to mitigate memory overhead. Therefore, to estimate the energy consumption of the \ac{SNN}, we initially train the network using quantized synaptic weights derived from device modeling. The effectiveness of \ac{DNN} stems from optimizing learning parameters through backpropagation and gradient descent methods. However, due to the discrete nature of spikes, direct implementation of gradient descent is impractical. To facilitate gradient-based learning in \acp{SNN}, the surrogate gradient method~\cite{neftci2019surrogate} is employed, utilizing a surrogate function whose derivative approximates a spike during backpropagation. Input non-binary images from the dataset are transformed into binary spikes using the Poisson encoder. Following successful network training, inference tasks are conducted to assess network accuracy and generate raster plots, which identify the indices of spiking neurons across different layers. Subsequently, knowledge of the spiking neuron indices enables the determination of synaptic weights connecting neurons from different layers within the trained model. Through circuit simulation, current and corresponding energy values can be derived for all spiking neurons and connecting synapses.

	\section{Spiking Neural Network simulation}\label{Sec:snn_simlation}
To conduct network-level simulations, we adopt a neural network topology comprising 784$\times$128$\times$10 neurons. The functionalities of neurons and synapses in the \ac{SNN} are emulated through our proposed~\cite{das2023bilayer} Leaky Integrate-and-Fire (LIF) neurons and 3-bit synapses, respectively. The simulation of the \ac{SNN} is performed using SpikingJelly~\cite{fang2023spikingjelly}, a framework based on PyTorch~\cite{paszke2019pytorch}. Our assessment of the \ac{SNN}'s performance is carried out using the \ac{MNIST} handwritten digit dataset, which comprises 60,000 training and 10,000 testing images, each of size 28$\times$28 pixels. To transform input pixel values into spikes, we employ a Poisson encoder, generating firing probabilities based on the Poisson distribution.
\begin{figure}[!h]
	\includegraphics[scale=0.5]{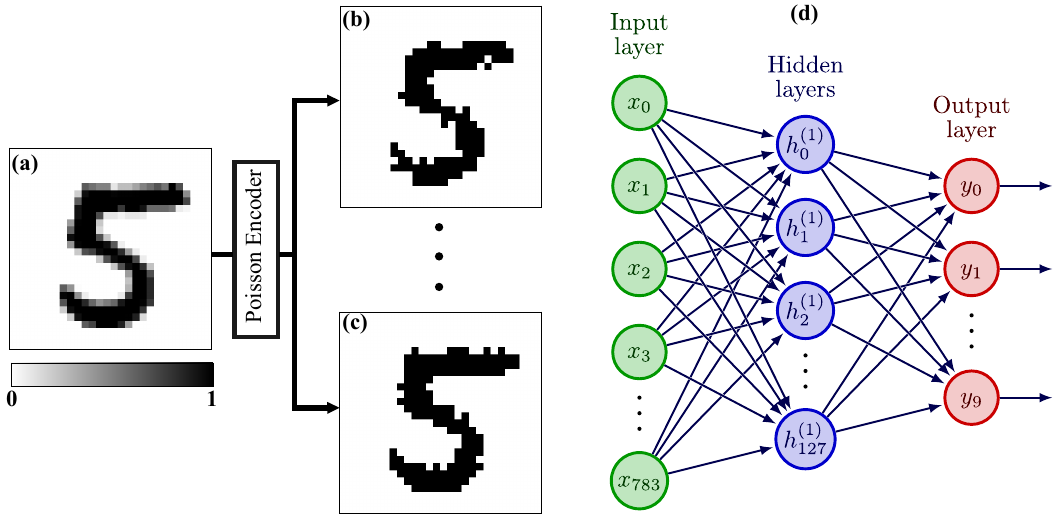}
	\caption{(a) \ac{MNIST} image for digit `5', (b), (c) Converted image using Poisson encoder. (d) \ac{SNN} topology of 784$\times$128$\times$10 neurons.}
	\label{Fig:SNN_image}
\end{figure}
Fig.~\ref{Fig:SNN_image}(a) shows a 28$\times$28 pixel image of digit `5', with a normalized pixel value ($\in \left[0,1\right] $). Utilizing the Poisson encoder, each image undergoes a transformation into $n$ distinct images, maintaining the same pixel count, with pixel values ($\in \left\lbrace 0,1\right\rbrace $) derived from a Poisson distribution. Figures~\ref{Fig:SNN_image}(b) and (c) depict two examples of such encoded images. Subsequently, these encoded images are converted into column vectors and introduced into the input layer of the \ac{SNN}, illustrated in Figure~\ref{Fig:SNN_image}(d). The efficacy of deep learning models primarily stems from their training methods based on gradient descent and backpropagation. However, due to the discrete nature of spikes, the outputs are non-differentiable. Consequently, conventional backpropagation cannot be employed for training the \ac{SNN}. To incorporate the principles of backpropagation into \acp{SNN}, we adopt the surrogate gradient method \cite{neftci2019surrogate}. In this approach, the spike-generating discontinuous Heaviside function is replaced by a continuous function (e.g., sigmoid)~\cite{fang2023spikingjelly}, facilitating the computation of derivatives of errors during the backward pass.
\begin{figure}[!h]
    \centering
    \includegraphics[scale=0.45]{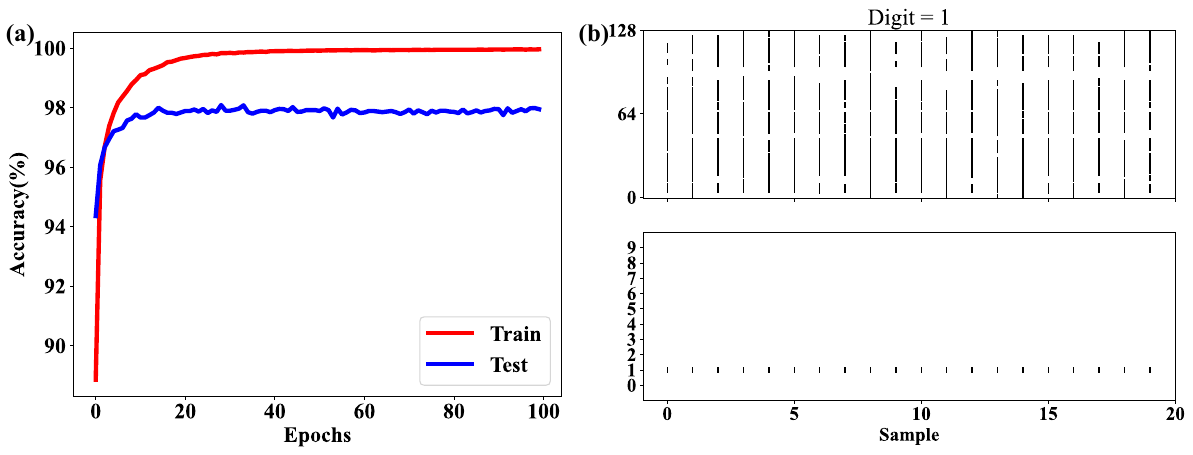}
    \caption{(a) Variation of training and test accuracies with the epoch. (b) Raster plot for first 20 test samples of digit `1', showing the firing neuron index of the hidden layer (top panel) and output layer (bottom panel).}
    \label{Fig:Acc_Raster}
\end{figure}

The training of the \ac{SNN} is executed using the Adam optimizer~\cite{kingma2014adam} with a batch size of 64 over 100 epochs. In Figure~\ref{Fig:Acc_Raster}(a), the training accuracy evolution across epochs is displayed. Simulation results reveal that the training accuracy surpasses 99\% after 10 epochs, reaching a maximum accuracy of 99.96\%, while the test accuracy stabilizes at approximately 98\% after 20 epochs. For the calculation of energy consumption during inference, the number of firing neurons is crucial, and this information can be extracted from the raster plot. We select 20 samples from the test dataset for each digit class, input the Poisson-encoded images into our trained network, and plot the indices of firing neurons for both the hidden and output layers. Figure~\ref{Fig:Acc_Raster}(b) illustrates a representative raster plot for the digit `1', with the top and bottom panels depicting the indices for the hidden and output layers across 20 samples, respectively. Upon obtaining the firing neuron indices for both layers, the total energy can be computed by multiplying the energy consumption of individual neurons and synapses. The energy consumption of these components can be determined through circuit-level simulations, a topic that will be elucidated in the subsequent section.
	
	\section{Device concept}\label{Sec:Device}
The magnetic skyrmion, a distinctive magnetic texture nucleated in a thin film of \ac{FM} grown on a \ac{HM}~\cite{woo2016observation,ma2016interfacial}, holds significant potential for energy-efficient computing due to its small size and ultralow depinning current density~\cite{yu2012skyrmion}. However, the design of compact devices encounters challenges arising from the Magnus force generated by the skyrmion Hall effect, affecting skyrmion motion within the monolayer \ac{FM}. To address this issue, a bilayer device is proposed~\cite{zhang2016magnetic}, wherein two \ac{FM} layers with opposite magnetization and coupled by an \ac{AFM} coupling, are separated by a spacer layer. It has been demonstrated that by establishing appropriate \ac{AFM} coupling between the two \ac{FM} layers, the Magnus force can be completely nullified, enabling straight-line motion of skyrmions. In our approach, we emulate the membrane potential using the position of the skyrmion in the nanotrack of the bilayer device. By introducing an anisotropy gradient along the length of the device, we incorporate a leak mechanism. During the presence (absence) of an incoming spike, the skyrmion exhibits forward (backward) movement, resembling integration (leak) behavior. Upon reaching a designated position under a detecting \ac{MTJ} at the right side of the nanotrack, the resistance of the corresponding \ac{MTJ} changes, generating an output spike followed by a reset mechanism that returns the skyrmion to its initial position. Additionally, synapse weights are emulated by incorporating multiple skyrmions in a single device. In a prior work~\cite{das2023bilayer}, Das et.al. demonstrated the emulation of neuron and synapse functionality in bilayer skyrmion devices. The results obtained from this study are utilized for the \ac{SNN} simulation. Interested readers are encouraged to refer to Ref~\cite{das2023bilayer} for a comprehensive understanding of the device and its operation.

\begin{figure}[t!]
	\includegraphics[scale=0.3]{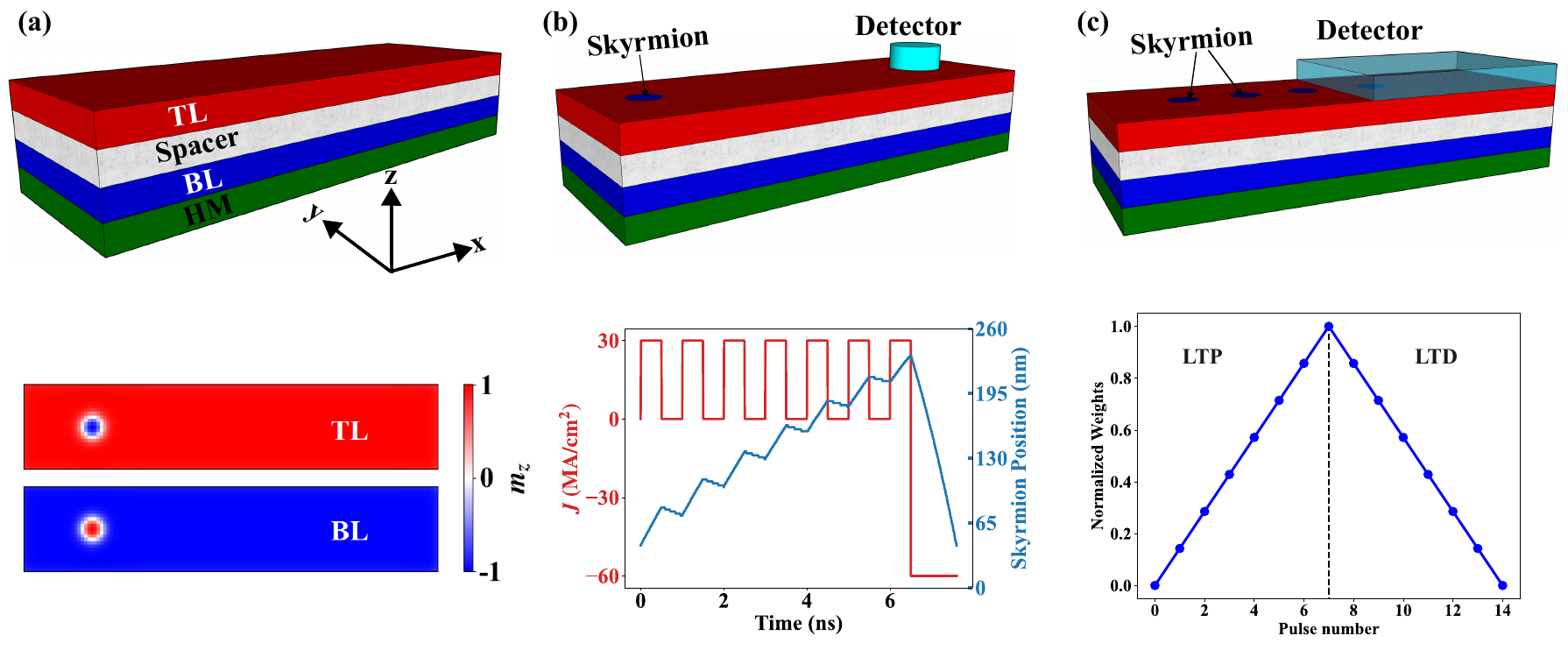}
	\caption{(a) (Top) Schematic 3d-view of the proposed bilayer device, (bottom) two-dimensional color plot of $z$-component ($m_z$) of magnetization on the $x\mbox{-}y$ plane for TL and BL. (b) (Top) Schematic 3d-view of the proposed bilayer neuron along with a skyrmion shown at TL and a detector, (bottom) Time variation of input current density (red) and corresponding skyrmion position (blue) along the length of the nanotrack. (c) (Top) Schematic 3d-view of the proposed bilayer synapse device, (bottom) plot of normalized synaptic weight with pulse number for LTP and LTD. }
	\label{Fig:device}
\end{figure}
The top panel of Figure~\ref{Fig:device}(a) presents a schematic 3D-view of our proposed device, featuring two \ac{FM} layers, denoted as the \ac{TL} and \ac{BL}), with opposite magnetization, coupled through an antiferromagnetic (\ac{AFM}) exchange interaction.  In the bottom panel, the $z$-component magnetization, $m_z$, of both \ac{TL} and \ac{BL} is illustrated, depicting nucleated skyrmions in both layers. The color bar highlights the antiferromagnetic coupling between the magnetizations of these two layers. The bottom panel of Figure~\ref{Fig:device}(b) displays the input current density (in red) alongside the reset current variation with time. Simultaneously, the corresponding position of the skyrmion along the $x$-direction of the nanotrack is depicted in the same figure. The assumed pulse width of the input spike is 0.5 ns.
Figure~\ref{Fig:device}(c) exhibits the \ac{LTP} and \ac{LTD} operation, wherein synaptic weights increase and decrease, respectively, with the pulse numbers. Our proposed synapse device achieves 8 (3-bit) distinct states, as indicated by the normalized weight in the figure. These characteristics of the neuron and synapse are employed in our \ac{SNN} simulation.

	\section{Associated circuit for the operation}\label{Sec:circuit}
\begin{figure}[!htbp]
    \centering
    \includegraphics[scale=0.185]{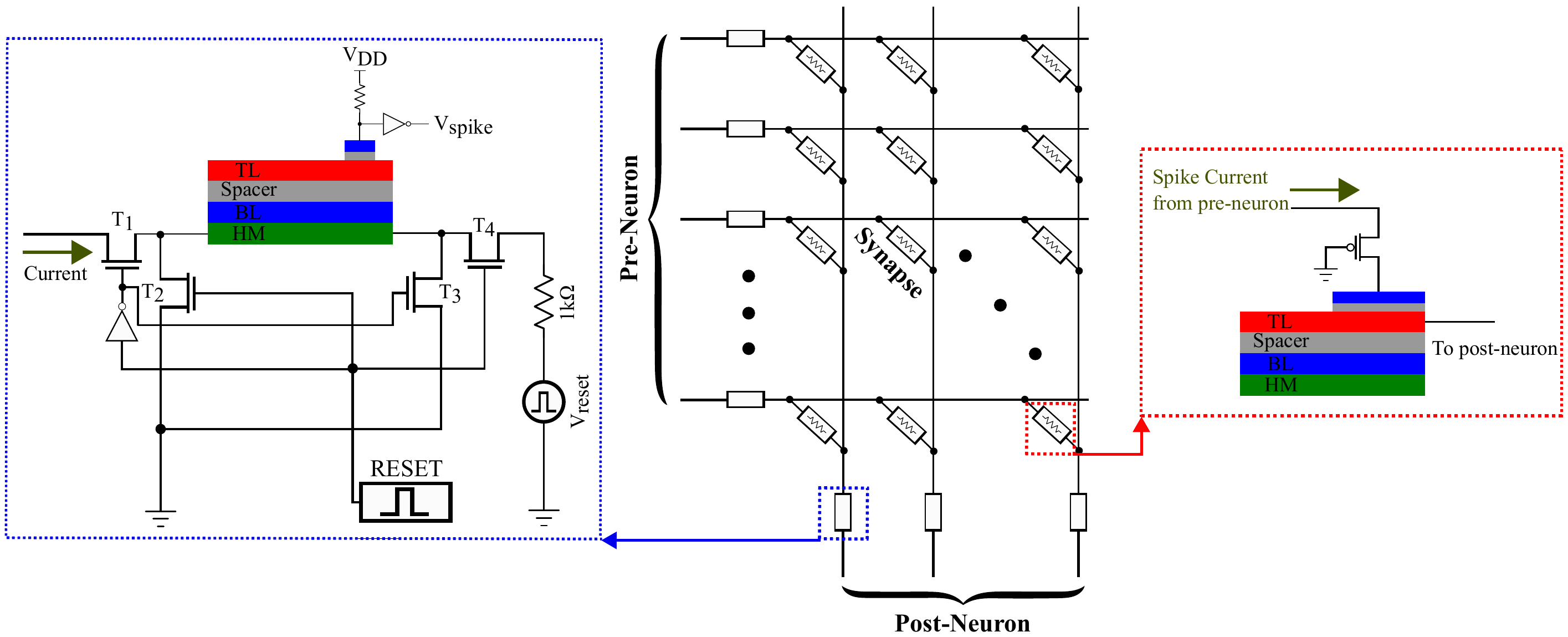}
    \caption{Schematic of the crossbar architecture, where pre-neurons are connected to the post-neurons via synapses. A zoomed-in view of the neuron (blue dashed box) and synapse (red dashed box) is shown along with the peripheral circuits.}
    \label{Fig:matrix}
\end{figure}
In the \ac{SNN}, connections exist between the pre-neurons of a layer and the post-neurons of the subsequent layer through synapses, illustrated in Fig.~\ref{Fig:matrix}. The figure provides a closer look at a neuron (enclosed in a blue dashed box) and a synapse (enclosed in a red dashed box), featuring our proposed devices and their corresponding circuits. Subsequently, we provide a comprehensive description of the circuit operation for both the neuron and synapse.

For our proposed neuron device, the lateral dimensions of the layers are set at 260~nm$\times$ 50~nm, each with a thickness of 1~nm. To induce the motion of the skyrmion along the nanotrack, a current is injected into the \ac{HM} layer along the $x$-direction, through $y\mbox{-}z$ plane. This results in a cross-sectional area ($A_{\mathrm{HM}}$) for the current injection of 50~$\mathrm{nm^2}$. Utilizing the resistivity of the HM, denoted as $\rho_{\mathrm{HM}}$=100 $\mu\Omega\mbox{-}\mathrm{cm}$ \cite{nguyen2016spin}, the resistance of the HM ($R_{\mathrm{HM}}$) can be calculated using the formula $R_{\mathrm{HM}}=\rho_{\mathrm{HM}} L_{\mathrm{HM}}/A_{\mathrm{HM}} = 5.2~\mathrm{k}\Omega$, where $L_{\mathrm{HM}}$ represents the length of the neuron device. Assuming a pulse current density magnitude of 30MA/$\mathrm{cm^2}$ during the integration phase, a current of 15 $\mathrm{\mu A}$ flows through the HM layer during an active pulse with a pulse width of 0.5 ns. For the reset operation, the current density is assumed to be 60MA/$\mathrm{cm^2}$ (corresponding to 30 $\mathrm{\mu A}$) for a duration of 1.1 ns.

\begin{figure}[t!]
	\includegraphics[scale=0.55]{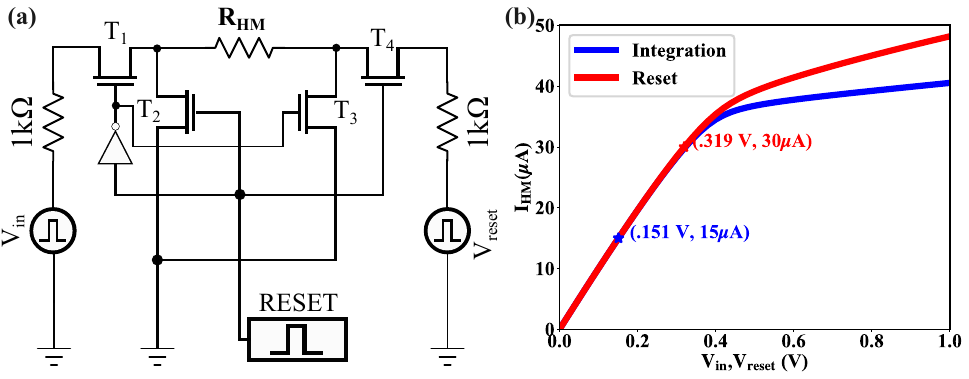}
	\caption{(a) Circuit diagram of neuronal operation. (b) Plot of $\mathrm{I_{HM}}$ with voltage sources $\mathrm{V_{spike}}$, and $\mathrm{V_{reset}}$ during `integration' and `reset' operation.}
	\label{Fig:Neuron_ckt}
\end{figure}
Fig.~\ref{Fig:Neuron_ckt} illustrates the requisite circuitry for the operation of the neuron, with the neuron represented by $\mathrm{R_{HM}}$ to account for the current flow through the resistance of the heavy metal (\ac{HM}). For practical considerations, we introduce non-ideal voltage sources $\mathrm{V_{spike}}$ and $\mathrm{V_{reset}}$, each containing an internal resistance of 1~$\mathrm{k\Omega}$, as depicted in Fig.~\ref{Fig:Neuron_ckt}. The MOSFET transistors $\mathrm{T_1\mbox{-}T_4}$ are utilized for this circuit, employing the FreePDK45 model and simulated in the Cadence Virtuoso simulator. Additionally, two more transistors associated with the inverter are not shown in the figure. In the circuit, the RESET block generates a high signal of 1V during the refractory period, and during the firing of the neuron, it remains inactive, maintaining a low signal of 0V. The RESET block's output is linked to the gate terminals of $\mathrm{T_2}$ and $\mathrm{T_4}$ and is fed into an inverter. The gate terminals of $\mathrm{T_1}$ and $\mathrm{T_3}$ are connected to the output of this inverter. Consequently, when the RESET block is inactive, $\mathrm{T_2}$ and $\mathrm{T_4}$ are deactivated, while $\mathrm{T_1}$ and $\mathrm{T_3}$ are activated, allowing the input spike current to flow through $\mathrm{R_{HM}}$ via $\mathrm{T_1}$ and $\mathrm{T_3}$ during the integration process. Conversely, when the RESET block is activated, it turns on $\mathrm{T_2}$ and $\mathrm{T_4}$ and turns off $\mathrm{T_1}$ and $\mathrm{T_3}$, allowing the reset current to flow through $\mathrm{R_{HM}}$ via $\mathrm{T_2}$ and $\mathrm{T_4}$. To determine the required magnitudes of $\mathrm{V_{in}}$ and $\mathrm{V_{reset}}$ for achieving integration and reset currents, these voltages are varied in the Virtuoso simulator. The plot in Fig.~\ref{Fig:Neuron_ckt}(b) illustrates the current ($\mathrm{I_{HM}}$) flowing through $R_{\mathrm{HM}}$ in relation to the variation of $\mathrm{V_{spike}}$ and $\mathrm{V_{reset}}$. From this plot, values of 0.151 V and 0.319 V are identified for $\mathrm{V_{in}}$ and $\mathrm{V_{reset}}$, respectively, to achieve integration and reset currents of 15 $\mathrm{\mu A}$ and 30 $\mathrm{\mu A}$. Considering the pulse widths of integration (0.5~ns) and reset (1.1~ns), the corresponding energy consumptions are determined to be 1.135~fJ and 10.56~fJ, respectively. Examining Fig.~\ref{Fig:device}(b), it is noteworthy that 7 pulses are required to move the skyrmion from its initial position to the detector. Accounting for all pulses, the total energy consumption for a neuron during its firing and reset processes is calculated to be 18.505~fJ.

\begin{figure}[h!]
	\includegraphics[scale=0.75]{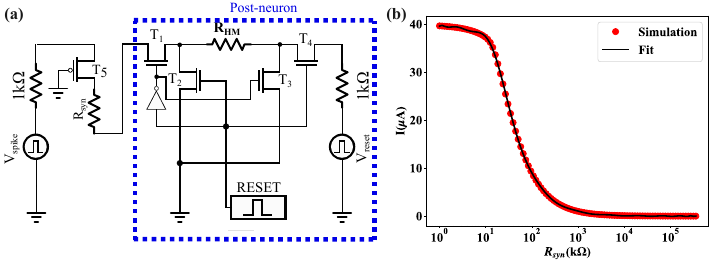}
	\caption{(a) Circuit diagram of synapse $\mathrm{R_{syn}}$ connecting the pre-neuron to the post-neuron. Simulation result (red dot) of variation of current passing through $\mathrm{R_{syn}}$, with the variation of its value, along with a fitted result (black) line.} 
	\label{Fig:Synapse_ckt}
\end{figure}

Spikes originating from the pre-neurons undergo modulation by the synapse weights, as illustrated in Fig.\ref{Fig:matrix}, where the different conductance states of the device represent these weights. In the bottom panel of Fig.\ref{Fig:device}(c), we depict the change in the proposed bilayer device's conductance on a normalized scale with respect to the pulse number. These discrete conductance states serve to represent the synaptic weights. Following the successful training of the Spiking Neural Network (\ac{SNN}), the synapse weights are acquired and can be converted into resistance. Let $w_{ij}$ denote the weight of the synapse connecting the $i^{th}$ pre-neuron to the $j^{th}$ post-neuron, resembling the conductance of the device. The corresponding resistance can be calculated as $R_{ij}=(1/|w_{ij}|)\cdot S$, where $S$ is a scaling factor to adjust the resistance to a desired range. Since the conductance states can be modulated with current, we represent the synaptic weight with a variable resistance $\mathrm{R_{syn}}$. The spike from the pre-neuron is fed to one end of $\mathrm{R_{syn}}$ via a PMOS transistor $\mathrm{T_5}$, and the other end of $\mathrm{R_{syn}}$ is connected to the post-neuron, as illustrated in Fig.\ref{Fig:Synapse_ckt}(a). Previously, the required current magnitude for integration was discussed to be 15$\mathrm{\mu A}$. Thus, it is crucial to ensure that a maximum of 15~$\mathrm{\mu A}$ current flows through the synapse at the minimum value of $\mathrm{R_{syn}}$.
As $\mathrm{R_{syn}}$ is variable, dependent on the synaptic conductance, it is varied within the resistance range obtained by the upper and lower limits of the synaptic weights for the trained network. The current flowing through $\mathrm{R_{syn}}$ due to this variation is plotted in Fig.~\ref{Fig:Synapse_ckt}(b), represented by the red dot. To obtain the current at any given resistance within this range, polynomial curve fitting is employed, as depicted by the black line.

\begin{figure}[h!]
    \includegraphics[scale=0.58]{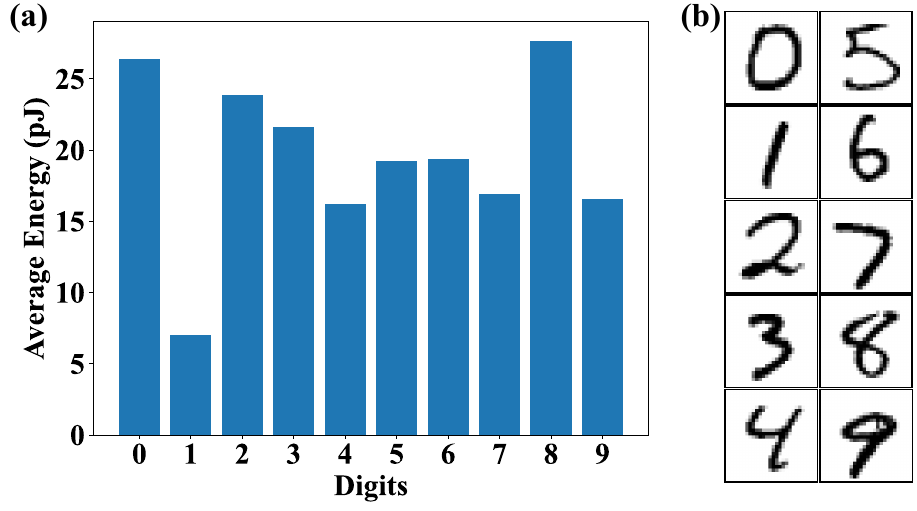}
    \caption{(a) Inference class-wise average energy consumption of the entire SNN. (b) Samples of each inference class}
    \label{Fig:MNIST_Energy}
\end{figure}

The accuracy of the network is evaluated using the test dataset, a process commonly referred to as \textit{inference}, following the successful training of the network. In the case of the \ac{SNN}, the identification of neurons that fired during the inference can be determined through a \textit{raster} plot. Figure~\ref{Fig:Acc_Raster}(b) presents such a plot for the digit `1' across 20 samples, with the top (bottom) panel showcasing the indices of the spiking neurons in the hidden (output) layer. Similar raster plots for other digits can also be obtained in a comparable manner, although they are not presented here. The number of neurons that spiked for 20 samples of individual classes, encompassing both hidden and output layers, is derived from the raster plots for each output digit class. \\
%\textbf{Energy calculation procedure}: 
During inference, the Poisson encoded spikes for each input image are presented in the input layer and the corresponding spikes at hidden and output layers can be obtained by the raster plot similar to Fig.~\ref{Fig:Acc_Raster}(b). Thus, if $\mathrm{E_N}$ is the energy required for the spiking process of a neuron, we can calculate the total energy contribution by multiplying $\mathrm{E_N}$ with $N_{neuron}$, the total number of spiking neurons (hidden and output layer). The \ac{SNN} was assumed to be fully connected, thus after the successful training, weight values of all the connecting synapses can be obtained. From the neuron's indices (obtained from the raster plot), the weight value of the corresponding synapses connecting interlayer neurons can also be obtained. Thus, from the fitting curve shown in Fig.~\ref{Fig:Synapse_ckt}(b), the current flowing through a synapse with a desired weight (resistance) value can be obtained. After getting the current value, the energy consumption by the synapse can be obtained by the current flowing through required $\mathrm{R_{syn}}$.

Taking into account the energy consumption values of the neurons (as discussed earlier) for both spiking neurons and connecting synapses, the average energy consumption for individual digit classes is calculated, as depicted in Figure~\ref{Fig:MNIST_Energy}(a). The figure reveals that the minimum and maximum energy consumption occurred for the digits `1' and `8', respectively. This observation may be attributed to the lower number of neurons that spiked during the inference of digit `1', while for digit `8', the number of firing neurons is notably higher. This disparity can be intuitively understood by considering the number of pixels associated with the input images, as illustrated in Figure~\ref{Fig:MNIST_Energy}(b). In this representation, it is evident that the number of pixels with non-zero values is minimal for digit `1', whereas for digit `8', this number is considerably higher.

	\section{Conclusion}\label{Sec:conclusion}
In conclusion, our research introduces a comprehensive framework for assessing hardware-level metrics in \acp{SNN} utilizing non-volatile memory devices, with a focus on the implementation of neurons and synapses using our proposed bilayer magnetic skyrmion device. The framework accommodates diverse non-volatile memory devices, extending its applicability beyond our proposed skyrmionic device to include other emerging technologies. Our implemented \ac{SNN} demonstrates robust performance, achieving training and testing accuracies of approximately 99\% and 98\%, respectively, on the \ac{MNIST} handwritten digit dataset. Notably, the average energy consumption during inference for individual digits is estimated in the order of pJ.
The results outlined in this study establish a pathway for a standardized evaluation methodology, providing researchers and practitioners in the field with the tools to make informed decisions when designing and implementing neuromorphic hardware. This research not only enhances our comprehension of the collaboration between SNNs and non-volatile memory but also lays the groundwork for future advancements in efficient and high-performance neuromorphic computing architectures.

    \section{Acknowledgments}\label{Sec:acknow}
This work is supported by the National Research Foundation, Prime Minister’s Office, Singapore, under its Competitive Research Program (NRF-CRP24-2020-0002), the Ministry of Education (Singapore) Tier 2 Academic Research Fund (MOE-T2EP50220-0012 and MOE-T2EP5-0221-0008), and the Agency for Science, Technology and Research (A*STAR) Programmatic Grant (A18A6b0057).
 
	%\pagebreak
	\bibliographystyle{IEEEtran}
	
	\bibliography{References}

\end{document}